\documentclass[11pt]{article}
\usepackage{graphicx}
\usepackage{amssymb}
\usepackage{dcolumn}
\usepackage{amsmath}
\usepackage{lscape}
\usepackage{longtable}

\begin{document}
\textwidth 16cm
\newcommand{\bd}{\begin{document}}
\newcommand{\ed}{\end{document}}
\newcommand{\bc}{\begin{center}}
\newcommand{\ec}{\end{center}}
\newcommand{\bfr}{\begin{flushright}}
\newcommand{\efr}{\end{flushright}}
\newcommand{\lt}{\left}
\newcommand{\rt}{\right}
\newcommand{\vs}{\vspace}
\newcommand{\hs}{\hspace}
\newcommand{\beq}{\begin{equation}}
\newcommand{\eeq}{\end{equation}}
\newcommand{\lb}{\linebreak}
\newcommand{\pb}{\pagebreak}
\newcommand{\mb}{\makebox}
\newcommand{\fb}{\framebox}
\newcommand{\mc}{\multicolumn}
\newcommand{\ben}{\begin{enumerate}}
\newcommand{\een}{\end{enumerate}}
\newcommand{\bit}{\begin{itemize}}
\newcommand{\eit}{\end{itemize}}
\newcommand{\ol}{\overline}
\newcommand{\un}{\underline}
\newcommand{\lefq}{\lefteqn}
\newcommand{\ba}{\begin{array}}
\newcommand{\ea}{\end{array}}
\newcommand{\beqa}{\begin{eqnarray}}
\newcommand{\eeqa}{\end{eqnarray}}
\newcommand{\beqas}{\begin{eqnarray*}}
\newcommand{\eeqas}{\end{eqnarray*}}
\newcommand{\bfg}{\begin{figure}}
\newcommand{\efg}{\end{figure}}
\newcommand{\bds}{\begin{displaymath}}
\newcommand{\eds}{\end{displaymath}}
\newcommand{\btb}{\begin{tabbing}}
\newcommand{\etb}{\end{tabbing}}
\bc {\huge $\mathcal{PT}$ Symmetric Hamiltonian Model and Exactly Solvable Potentials } \ec

\vs{0.5cm}

\bc
{\it \"Ozlem Ye\c{s}ilta\c{s}{\footnote {e-mail : yesiltas@gazi.edu.tr}\\
Department of Physics, Faculty of Science,
Gazi University,
06500 Ankara, Turkey\\
\vspace{.12cm}

}} \ec
\vs{0.5cm}
\begin{abstract}
 \noindent Searching for non-Hermitian (parity-time)$\mathcal{PT}$-symmetric Hamiltonians \cite{bender} with real spectra has been acquiring much interest for fourteen years. In this article, we have introduced a $\mathcal{PT}$ symmetric non-Hermitian Hamiltonian model  which is given as  $\hat{\mathcal{H}}=\omega (\hat{b}^{\dag}\hat{b}+\frac{1}{2})+  \alpha (\hat{b}^{2}-(\hat{b}^{\dag})^{2})$ where $\omega$ and $\alpha$ are real constants, $\hat{b}$ and $\hat{b^{\dag}}$ are first order differential operators. Moreover, Pseudo-Hermiticity that is a generalization of $\mathcal{PT}$ symmetry has been attracting a growing interest \cite{mos}. Because the Hamiltonian $\mathcal{H}$ is pseudo-Hermitian, we have obtained the Hermitian equivalent of $\mathcal{H}$ which is in Sturm- Liouville form leads to exactly solvable potential models which are effective screened potential and hyperbolic Rosen-Morse II potential. $\mathcal{H}$ is called pseudo-Hermitian, if there exists a Hermitian and invertible operator $\eta$ satisfying $\mathcal{H^{\dag}}=\eta \mathcal{H} \eta^{-1}$. For the Hermitian Hamiltonian $h$, one can write $h=\rho \mathcal{H} \rho^{-1}$ where $\rho=\sqrt{\eta}$ is unitary. Using this $\rho$ we have obtained a physical Hamiltonian $h$ for each case. Then, the Schr\"{o}dinger equation is solved exactly using Shape Invariance method of Supersymmetric Quantum Mechanics \cite{susy1}. Mapping function $\rho$ is obtained for each potential case.

\end{abstract}

\section{Introduction}
There is growing interest in the study of non-Hermitian Hamiltonian operators \cite{bender, mos,sinha, ahmed, bd} particularly those which possess $\mathcal{PT}$ symmetry. Ten years ago, pseudo-Hermiticity concept is introduced in a parallel development \cite{mos}. A Hermitian operator $\eta$ and a pseudo-Hermitian Hamiltonian $\mathcal{H}$ satisfy $\mathcal{H}^{\dag}=\eta \mathcal{H} \eta^{-1}$. We note that $\eta$ corresponds to $\mathcal{PC}$ in the case of $\mathcal{PT}$ symmetric Hamiltonians. It is also shown that a Hamiltonian is equivalent to a Hermitian Hamiltonian $h$ under a similarity transformation $\mathcal{H}=\rho^{-1} h \rho$ \cite{mos}. In this paper we wish to explore a pseudo-Hermitian Hamiltonian which is introduced by us and which is a special form of the Swanson Hamiltonian in \cite{swanson} and also in \cite{swanson2, yes}, a $\mathcal{PT}$ symmetric non-Hermitian model with two parameters,
\begin{equation}\label{4}
  \hat{\mathcal{H}}=\omega (\hat{b}^{\dag}\hat{b}+\frac{1}{2})+  \alpha (\hat{b}^{2}-(\hat{b}^{\dag})^{2})
\end{equation}
where $\hat{b}, \hat{b^{\dag}}$ are bosonic annihilation and creation operators, $\omega, \alpha$  are real constants and  $\dag$ is Hermitian adjoint, $\hat{b}$ is the annihilation operator given in a general form $ \hat{b}=A(x)\frac{d}{dx}+B(x)$ and $A(x)$, $B(x)$ are real functions. In terms of differential operators, our Hamiltonian operator is written as
\begin{equation}\label{6}
\begin{split}
  \hat{\mathcal{H}}&=-\omega A(x)^{2}\frac{d^{2}}{dx^{2}}+(4\alpha A(x)B(x)-2\omega A(x)A(x)^{'})\frac{d}{dx}\\-
  &(\omega-2 \alpha) A(x)B(x)^{'}-(\omega- 2 \alpha) A(x)^{'} B(x)+\omega B(x)^{2}-\alpha (A(x)A(x)^{''}+(A(x)^{'})^{2})+\frac{\omega}{2}.
  \end{split}
\end{equation}
We may write the eigenvalue equation as $\hat{\mathcal{H}} \psi=\varepsilon \psi$ and here, the pseudo-Hermitian Hamiltonian $\mathcal{H}$ can be mapped into a Hermitian operator form by using a mapping function $\rho$ where $  \rho(x)=e^{-\frac{2\alpha}{\omega}\int^{x} dy \frac{B(y)}{A(y)}}$.
Here we note that $h \psi=\varepsilon \psi$, $\psi=\rho^{-1}\xi$. So we can introduce operator $h$ which is Hermitian equivalent of $\mathcal{H}$ as
\begin{equation}\label{10}
 h= -\omega\frac{d}{dx}A(x)^{2}\frac{d}{dx}+U_{eff}(x)
\end{equation}
\begin{equation}\label{11}
  U_{eff}(x)=\frac{\omega}{2}-\omega(A(x)B(x))^{'}-\alpha \left((A^{'}(x))^{2}+A(x) A^{''}(x)\right)+(\omega+\frac{4\alpha^{2}}{\omega})B^{2}(x)
 \end{equation}
where the primes denote the derivatives. Then (\ref{10}) can be mapped into a Schr\"{o}dinger-like form by using $ \xi(x)=\frac{1}{A(x)} \Phi(x)$ which leads to $    -\Phi^{''}+\bar{U}_{eff}(x)\Phi=\frac{\varepsilon}{\omega A^{2}(x)} \Phi$. Hence, $\bar{U}_{eff}(x)$ becomes
\begin{equation}\label{14}
  \bar{U}_{eff}(x)=\frac{A^{''}(x)+1/2-(A(x)B(x))^{'}}{A(x)}-\frac{\alpha}{\omega}\frac{(A^{'}(x))^{2}+A^{''}(x)}{A(x)}+
  \frac{\omega^{2}+4\alpha}{\omega^{2}}  \frac{B^{2}(x)}{A^{2}(x)}.
\end{equation}
Let us take the mapping $\xi(x)=e^{\int dx \frac{4\alpha B(x)-\omega A(x)^{'})}{2\omega A(x)}}\chi(x)$, then eigenvalue equation for (\ref{10}) becomes
\begin{equation}\label{15}
\begin{split}
    -\omega A(x)^{2} \chi(x)^{''}-\omega A(x)A(x)^{'}\chi(x)^{'}+ [\frac{\omega}{2}-(\omega+2\alpha)(A(x)B(x))^{'}&+(\omega+\frac{8\alpha^{2}}{\omega})B^{2}(x)+\\ \omega/2
    +(\omega/4-\alpha)(A(x)^{'})^{2}+(\omega/2-\alpha)A(x)A^{''}(x)]\chi(x)=\varepsilon \chi(x).
    \end{split}
\end{equation}
If we use $y=\int^{x} \frac{dx}{A(x)}$ in (\ref{15}), we obtain
\begin{equation}\label{151}
\begin{split}
     -\omega\frac{d^{2}\chi}{dy^{2}}+[\frac{\omega}{2}-(\omega+2\alpha)(A(x)B(x))^{'}+ (\omega+\frac{8\alpha^{2}}{\omega})B^{2}(x)+ \omega/2
    +(\omega/4-\alpha)(A(x)^{'})^{2}&+ \\ (\omega/2-\alpha)A(x)A^{''}(x)]_{x\rightarrow y}\chi(y)=\varepsilon \chi(y)
\end{split}
\end{equation}

\section{Rosen-Morse II potential}
Let us choose $A(x)$ and $B(x)$ in these forms: $A(x)=a \cos h \mu x, ~~~~B(x)=-\beta_{1} A(x)$. These choices lead to
\begin{equation}\label{16}
    \bar{U}_{eff}(x)-\frac{\varepsilon}{\omega A^{2}(x)}=\left(\frac{(\omega-2\epsilon)}{a^{2}\omega}-\frac{(2\omega-\alpha)\mu^{2}}{\omega}\right)\sec h^{2}(\mu x)+2\beta_{1}\mu \tanh (\mu x)+\frac{2\omega(\omega-\alpha)\mu^{2}+4\alpha^{2}\beta_{1}^{2}}{\omega^{2}}.
\end{equation}
The eigenvalue equation becomes
\begin{equation}\label{e.v.}
  \Phi^{''}(x)+\left(\frac{\varepsilon}{\omega A^{2}(x)}-\bar{U}_{eff}(x)\right)\Phi(x)=E\Phi(x)
\end{equation}
where $E=2(\omega^{2} \mu^{2}+2\alpha^{2}\beta_{1}^{2})/\omega^{2}$. If we put the super-potential $W(x)$ in the form of $W(x)=A+B \tanh (\mu x)$, then we have the ground-state wave-function $\Phi_{0})(x)=e^{-Ax}(\cosh (\mu x))^{-B/\mu}$ where $B>0$ and $|A|<B$ because of the boundary conditions $x\rightarrow \pm \infty$, $\Phi_{0}(x)\rightarrow 0$. Now we can give the partner potentials $V_{-}(x)$ and $V_{+}(x)$ as
\begin{eqnarray}\label{a}
  V_{-}(x) &=& A^{2}+B^{2}+2\beta_{1}\mu\tanh (\mu x)-B(B+\mu)\sec h^{2}(\mu x)=  \bar{U}_{eff}(x)-\frac{\varepsilon}{\omega A^{2}(x)}-E_{0} \label{b} \\
  V_{+}(x) &=& A^{2}+B^{2}+2\beta_{1}\mu\tanh (\mu x)-B(B-\mu)\sec h^{2}(\mu x) \label{c}
\end{eqnarray}
 and we match (\ref{16}) and (\ref{b}), then we have
\begin{eqnarray}\label{a}
  2AB &=& 2\beta_{1}\mu \\
  B &=& -\frac{\mu}{2}+\frac{1}{2\sqrt{\omega}}\sqrt{8\epsilon-4\omega+a^{2}\mu^{2}(9\omega-4\alpha)} \label{c} \\
  E_{0} &=& -\left(B^{2}+\left(\frac{\beta_{1}\mu}{B}\right)^{2}\right).
\end{eqnarray}
Shape invariance relationship between the partner potentials is known as $V_{+}(x,a_{0})=V_{-}(x,a_{1})+R(a_{1})$ where $R(a_{1})$ is the reminder independent of $x$ and $a_{0}=B$, $a_{1}=B-\mu$.Because the energy spectrum is given by $E_{0}^{-}=0$, $E_{n}^{-}=\sum^{n}_{k=1} R(a_{k})$, we can obtain the spectrum of (\ref{16}) as
\begin{equation}\label{17}
  E^{-}_{n}=-\left(\frac{\beta_{1}\mu}{B-n\mu}\right)^{2}-(B-n\mu)^{2}-\frac{2\alpha \mu^{2}}{\omega}, ~~~~n=0, 1, 2,...
\end{equation}
To find corresponding eigenfunctions of the system, we will use $E^{-}_{n}$ in eigenvalue equation (\ref{e.v.}) then we obtain un-normalized $\Phi_{n}$,
\begin{equation}\label{18}
  \Phi_{n}(x)=B_{n}(1+\tanh (\mu x))^{-r} (1-\tanh (\mu x))^{-s} P^{(-2r, 2s)}_{n}(-\tanh (\mu x))
\end{equation}
where $B_{n}$ is the normalization constant, $r=\frac{1}{2}(n+\nu-\frac{\beta_{1}}{\mu}\frac{1}{n+\nu})$, $s=\frac{1}{2}(n+\nu+\frac{\beta_{1}}{\mu}\frac{1}{n+\nu})$, $\nu=B/\mu$ and $P^{(-2r, 2s)}_{n}(-\tanh (\mu x))$ stands for Jacobi polynomials. For the normalizability, we may give the $\rho=e^{\frac{2\alpha \beta_{1}x}{\omega}}$ used in Hermitian inner product  $\ll \Phi | \Phi \gg = <\Phi | \rho^{2} \Phi > $. The boundary conditions require $r>0, s>0$.
\section{Effective Screened potential}
Now we can choose $A(x)$ and $B(x)$ in these forms: $A(x)=a e^{-\delta x+\tau}-q, ~~~~B(x)=-b A(x)$. These choices lead to
\begin{equation}\label{18}
    \bar{U}_{eff}(x)-\frac{\varepsilon}{\omega A^{2}(x)}=-2b\delta+b^{2}\frac{\omega^{2}+4\alpha^{2}}{\omega^{2}}-\frac{a \alpha \delta^{2}}{\omega}
    \frac{e^{-\delta x+\tau}}{ae^{-\delta x+\tau}-q}+\left(2a^{2}\delta^{2}-a^{2}\alpha \frac{\delta^{2}}{\omega}\right)\frac{e^{-2\delta x+2\tau}}{(ae^{-\delta x+\tau}-q)^{2}}
\end{equation}
where we use $\omega=\epsilon/2$ and $\frac{a\delta}{2b}=q$ parameter restrictions. If we give the super-potential $W(x)=A+B\frac{e^{-\delta x}+\tau}{e^{-\delta x+\tau}-q}$, then we have the ground-state function in the form of $\Phi_{0}(x)=e^{-(A+B/a)x}(-ae^{\tau}+e^{\delta x}q)$. The boundary conditions require $(A+B/a) >0$. If we consider the partner potentials $V_{\mp}(x)$ for this case,
\begin{equation}\label{19}
    V_{\mp}(x)=A^{2}+(B^{2}\mp a\delta B)\frac{e^{-2\delta x+2\tau}}{(ae^{-\delta x+\tau}-q)^{2}}+
    B(2A\pm \delta)\frac{e^{-\delta x+\tau}}{ae^{-\delta x+\tau}-q}
\end{equation}
then we have,
\begin{eqnarray}
  B &=& \frac{a\delta}{2}+\frac{a\delta}{2}\sqrt{9-\frac{4\alpha}{\omega}} \\
  A &=& -\frac{\delta}{2}-\frac{\alpha \delta}{\omega\left(1+\sqrt{9-\frac{4\alpha}{\omega}}\right)} \\
  E_{0} &=& -A^{2}, ~~~~E=2b\delta- b^{2}\frac{\omega^{2}+4\alpha^{2}}{\omega^{2}}.
\end{eqnarray}
It is noted that $(a_{0}, b_{0})=(A, B)$ and $(a_{1}, b_{1})=(A-\alpha, B+\delta a)$. Then, we have obtained the energy spectrum as
\begin{equation}\label{20}
    E^{-}_{n}=-\left(\delta \sqrt{9-\frac{4\alpha}{\omega}}+\frac{\alpha \delta/\omega}{n+\frac{1}{2}+\frac{1}{2}\sqrt{9-\frac{4\alpha}{\omega}}}-
    \delta\left(n+\frac{1}{2}+\frac{1}{2}\sqrt{9-\frac{4\alpha}{\omega}}\right)\right)^{2}, ~~~~n=0,1,2,...
\end{equation}
Following the same procedure shown in the last section, the corresponding wave-functions are obtained as
\begin{equation}\label{21}
    \Phi_{n}(x)=B_{n}s^{E}(-q+as)^{\frac{q+\sqrt{q^{2}-4\gamma}}{2a}} P^{(2E, \frac{\sqrt{q^{2}-4\gamma}}{a})}_{n}(-q+2as)
\end{equation}
where $B_{n}$ is the normalization constant and $s=e^{-\delta x+\tau}$. The mapping function $\rho=e^{\frac{2\alpha b x}{\omega}}$ can be given for this case.
\section{Conclusions}
We have found real energy spectrum and corresponding eigenfunctions of a non-Hermitian, parity-time-symmetric Hamiltonian. We have generated two specific potentials, hyperbolic Rosen Morse and effective screened potentials namely, we have seen that real energy is guaranteed if inside of the square root is positive in (\ref{c}) that leads to restrictions as $-\sqrt{\frac{8\epsilon-4\omega}{a^{2}(4\alpha-9\omega)}}<\mu<\sqrt{\frac{8\epsilon-4\omega}{a^{2}(4\alpha-9\omega)}}$. We have derived effective screened potential under parameter restrictions, real spectrum is guaranteed if $9\omega-4\alpha >0$. Shape invariance technique is applied to obtain the solutions, the mapping functions that transforms non-Hermitian operator into a Hermitian operator are written for each case. In figure 1, for the first potential example, it is shown that when $\mu$ gets smaller, singularity appears for the large $n$ quantum numbers which means small values of $\mu$ leads to an approach to a classical system.



\end{document}